# Cluster-based MDS Algorithm for Nodes Localization in Wireless Sensor Networks with Irregular Topologies

Biljana Stojkoska
Faculty of Electrical Engineering and Information Technologies,
Karposh 2, b.b  Skopje
+38970435040
biles@feit.ukim.edu.mk

Danco Davcev
Faculty of Electrical Engineering and Information Technologies,
Karposh 2, b.b  Skopje
+38923099155
etfdav@feit.ukim.edu.mk

Andrea Kulakov
Faculty of Electrical Engineering and Information Technologies,
Karposh 2, b.b  Skopje
+38923099157
kulak@feit.ukim.edu.mk

## ABSTRACT
Nodes localization in Wireless Sensor Networks (WSN) has arisen as a very challenging problem in the research community. Most of the applications for WSN are not useful without a priori known nodes positions. One solution to the problem is by adding GPS receivers to each node. Since this is an expensive approach and inapplicable for indoor environments, we need to find an alternative intelligent mechanism for determining nodes location. In this paper, we propose our cluster-based approach of multidimensional scaling (MDS) technique. Our initial experiments show that our algorithm outperforms MDS-MAP[8], particularly for irregular topologies in terms of accuracy.

## Categories and Subject Descriptors
C.2.1 [**Computer-Communication Networks**]: Network Architecture and Design - Centralized networks, Network topology, Wireless communication

## 1. INTRODUCTION
In the recent years, there has been great advancement in the wireless sensor computing technology [2]. The sensors are organized in a network and communicate by exchanging information using radio modules. After taking samples from the environment they sense (light level, air temperature, humidity etc.), they can process data or exchange it. The processed packets are sent to the sink node. All data is not directly sent to the sink as sensors nodes have a short range of radio communication while being deployed in a vast region. There are a lot of multi-hops routing protocols that offer optimal communication cost. Each sensor sends data to its closest neighbor responsible for retransmitting the packets [3].

The problem of nodes localization appears in a variety of wireless sensor network (WSN) applications. The information gathered from the network can often be useless if not matched with the location where it is sensed. Finding the exact physical location is a crucial issue for continual network operation and WSN management. Many different techniques have been proposed for solving this problem, however, since most of them fail to perform well on irregular topologies, this problem remains a challenge.

Multidimensional Scaling (MDS) is a set of analytical techniques that have been used for many years in disciplines such as mathematical psychology, economics and marketing research. It is a suitable method used for reducing the dimensionality of the data, showing multidimensional data as e-points in two or three dimensional space [5]. This technique can also be used in WSN where only distances between nodes are known. The main advantage in using MDS is that it performs very accurate position estimation even when there are no anchor nodes. The distance measurement between each node is used as an input data. Although this method outperforms many others, it still provides poor results for irregular topologies. Since MDS is a centralized technique, all measurements are collected at the sink node where further processing is done. Other drawback of MDS is time complexity. Sending all distance measurements to the sink node may result in unaccepted occupation of the bandwidth which will reduce the efficiency of the system.

In this paper, we propose a modification of classical Multidimensional Scaling technique in order to improve its performance. The general idea of our cluster-based MDS algorithm is to generate local maps within each cluster. Local maps are then merged into a global one in order to estimate the position of the nodes. If given sufficient number of anchor nodes (nodes with a prior known location), this global map can be transformed into an absolute map.

The rest of the paper is organized as follows: In the second section, the relevant work related to the present localization techniques is discussed. The third section gives a detailed explanation of our cluster-based MDS algorithm (CB-MDS). The fourth section gives the results provided from the experiments. Finally, we conclude this paper in section five.

## 2. RELATED WORK

Many research groups have investigated different techniques for nodes localization in WSN. Most of the techniques proposed within the last years can be basically divided into two categories: range-based and range-free methods.

Range-free methods are also known as "hop-based" methods. They use hop or connectivity information for discovering nodes location.

The category of range-based methods estimates the distance between the neighboring nodes using different signal measurement techniques. RSSI (Receive Signal Strength Indicator) is the most common technique used since it doesn't require any additional hardware. Other popular techniques are ToA (Time of Arrival), AoA (Angle of Arrival) and TDoA (Time Difference of Arrival). They all require sensors equipped with powerful CPUs and appear as an expensive solution.

Multidimensional scaling (MDS) based algorithms are range-based sensor localization algorithms. There are different versions of MDS for nodes localization, the most popular is MDS-MAP, proposed by Yi Shang and Wheeler Ruml [8]. They showed that MDS-MAP outperforms other techniques, especially when applied on density networks. But this centralized approach gives significant errors for irregular topologies, such as C-shaped topology. Other approaches based on MDS exist, but they are complex and thus computationally dependent. Such an example is MDS-MAP(P) [9], which is a modification of the MDS-MAP based on a decentralized approach. It shows better results than MDS-MAP, but requires intensive computational resources at each node. MDS-MAP(P) computes local maps from each node in the network and then merges local maps into a global map.

The algorithm proposed in this paper is based on MDS and shows good results for irregular topologies in terms of accuracy and performance. Our cluster-based MDS is similar to MDS-MAP(P), but it calculates only local maps for each cluster instead of each node. Thus, only one node in the cluster is doing the computation. Additionally, for irregular C-shaped and H-shaped topologies, our algorithm gives more accurate prediction of the sensor locations than MDS-MAP. Although there has been proposed variations of cluster-based approaches (for example, see [1]), we used the metric proposed in [8], in order to provide comparable results with the reference MDS-MAP algorithm.

## 3. CLUSTER-BASED MDS

In this section we will explain in detail our cluster-based MDS algorithm for nodes localization within WSN.

Since radio signals are omni-directional, only nodes within certain radio range R can communicate with each other. If two nodes are within each others transmission range they are called neighbors. Further, we made the following assumptions:

- There is a path between every pair of nodes.
- Nodes that belong to the same cluster are deployed in a small geographical area. In other words, each cluster consists of nodes in close proximity to each other.
- Each node uses RSSI method for distance estimation.
- RSSI provide accurate neighboring sensor distance estimation.

Our cluster-based MDS algorithm is divided into four phases described below:

1. Initial clustering
2. Cluster extension
3. Local map construction
4. Local map merging

In the initial clustering phase, the network is divided into subsets called clusters. There are a lot of algorithms for nodes clustering in WSN [4][6]. In this paper, node clustering is not subject of interest, so it is assumed that network is already clustered by clustering algorithm. Each cluster consists of several neighbors nodes grouped together. In each cluster one representative node is chosen to be a cluster-head. Other nodes in the cluster are called members of the cluster. Clusters are disjunctive sets, allowing each node in the network to belong to only one cluster.

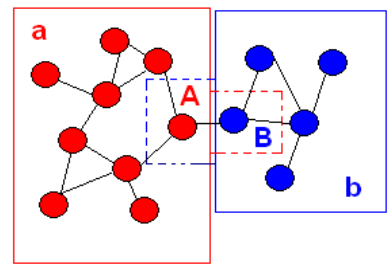

**Figure 1. Cluster extension**

In the second phase, clusters are extended by adding additional nodes from neighboring clusters. If node A from the cluster a has one hop neighbor node B that belongs to other cluster b, then node B is added to the cluster a, and node A is added to the cluster b (Figure 1). Thus nodes A and B become gateways. Gateways act as cluster members at the same time. Each gateway node participates in at least two clusters.

In the third phase, MDS-MAP algorithm generates local maps of each cluster. This phase has three sub-steps:

3.1. All member nodes send their measured distances to the cluster-head.

3.2. Cluster-head computes the shortest paths between all pairs of nodes that belong to the cluster and construct the distance matrix.

3.3 Cluster-head apply classical MDS to the distance matrix. The output of MDS is a local map that consists of relative positions of the cluster members.

Members of each cluster send their neighboring distance measurements to the cluster-head. After receiving the estimated distances, the cluster-head creates the distance matrix. If there are no measured distances between some of the nodes in the cluster, the cluster-head calculates these distances using Dijkstra's shortest path algorithm. When the distance matrix is filled with appropriate values (the true distances or shortest path distances), the cluster-head applies multidimensional scaling technique to the matrix. The output from MDS-MAP gives the coordinates of sensors within that cluster. At the end of the third phase, each cluster-head has the coordinates of its members. If the network consists of n clusters, then there are n local maps, each in different coordinate system.

In the forth phase merging of the local maps is provided using gateways. At least two neighboring clusters have 3 or more common nodes and they are called common-gateways. In both local maps common-gateways have different coordinates. One of these two neighboring clusters is known as a master and is denoted as Cm, while the other is the slave and is denoted with Cs. All member node coordinates that belong to the master are declared to be the correct positions. They are also known as local anchors. Common-gateways coordinates form Cm are also declared as local anchors and their coordinates are considered to be correct positions. At the same time, their coordinates in Cs are considered to be relative positions. Then the best linear transformation is applied to all nodes in Cs and their relative positions are aligned to the correct positions. The alignment consists of shift, rotation and reflection of coordinates [7]. All nodes in Cm will retain their positions, while all nodes in Cs will have new positions obtained after the alignment. Then all Cs members will become local anchors. It is worth to mention here that this phase can be done either iterative or in parallel which enables better overall performance. This process continues until the global map is generated.

If classical MDS is applied on distance matrix, nodes positions are estimated without an error. Since only the distances between neighboring nodes are known, Dijksta algorithm is used to find the shortest path between each node. This approximation produces an error, i.e. the correct positions usually differ from the predicted ones. This error is bigger when the nodes are in multi-hop communication range and that is the main reason why MDS-MAP reports very poor results on irregular topologies. In cluster-based MDS, multidimensional scaling is applied to each cluster. Nodes in the cluster are close to each other and shortest path distance estimation error is very small. Thus cluster-based MDS is expected to give better performance than MDS-MAP for irregular topologies.

## 4. SIMULATION RESULTS

We simulated CB-MDS algorithm on different network topologies with Matlab. Our work was mainly focused on irregular topologies (both grid and random), but we also considered other network properties like number of clusters, number of anchors and average connectivity.

- We consider grid and random deployment of the nodes for C-shaped, L-shaped and H-shaped topologies.
- Each node location was discovered with MDS-MAP algorithm and CB-MDS algorithm (using 5, 7, 10 and 15 clusters).
- Different number of global anchors (3, 4, 6 and 10) were used for generating the absolute map.
- Radio range changes leads to different average connectivity (average number of neighbors).

Thus 600 different networks were simulated (6 x 5 x 4 x 5).

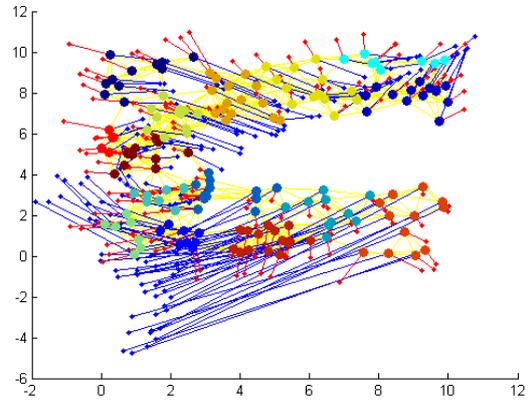

**a) R=1.5r**

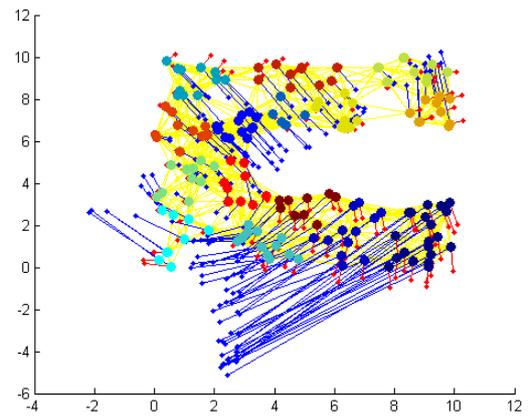

**b) R=2.0r**

**Figure 2. 161 nodes randomly deployed in C-shaped topology (15 clusters, 4 anchors)**

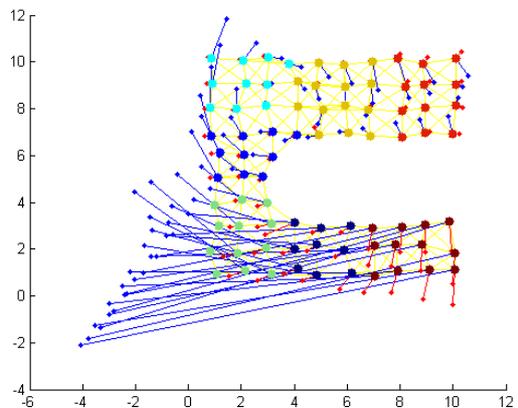

**a) R=1.8r**

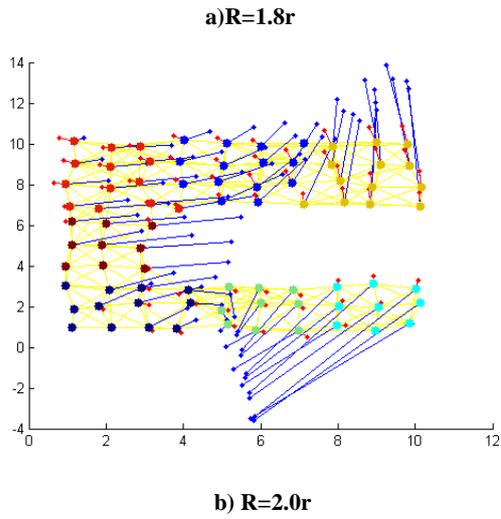

**b) R=2.0r**

**Figure 3. 79 nodes form C-shaped grid topology (7 clusters, 4 anchors)**

Figure 2 to Figure 5 show the results of MDS-MAP and CB-MDS on different topologies. The coloured circles correspond to the absolute position of the nodes for each topology. Nodes with the same colour belong to one cluster. The yellow lines represent the edges between neighboring nodes. Blue lines are the distance between the absolute and the estimated position when using MDS-MAP. Red lines are the distance between the absolute and the estimated position when using our cluster-based MDS algorithm. The error is larger if the lines are longer. It can be noted that cluster-based MDS gives better estimation results because the blue lines are much longer.

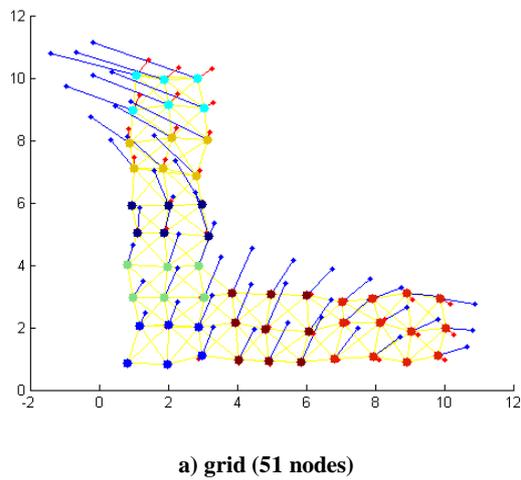

**a) grid (51 nodes)**

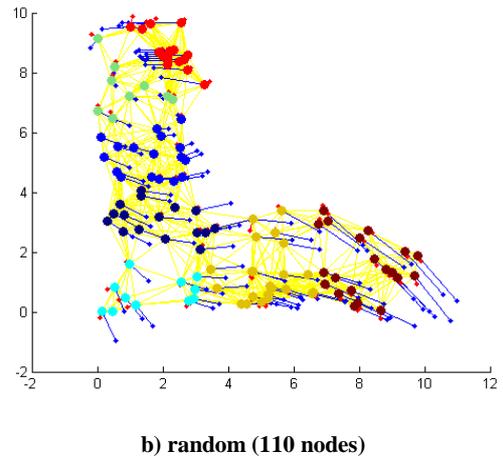

**b) random (110 nodes)**

**Figure 4. L-shaped topology (7 clusters, 4 anchors, R=1.8r)**

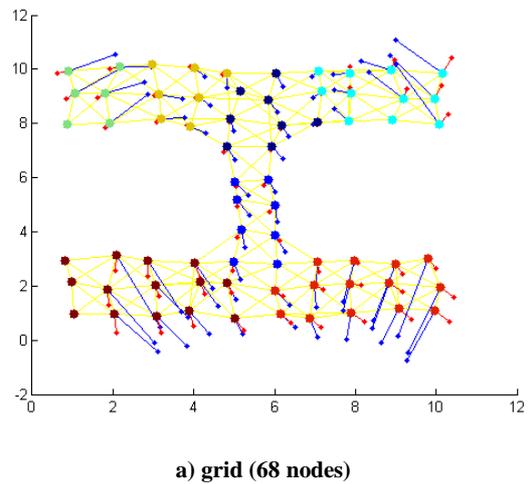

**a) grid (68 nodes)**

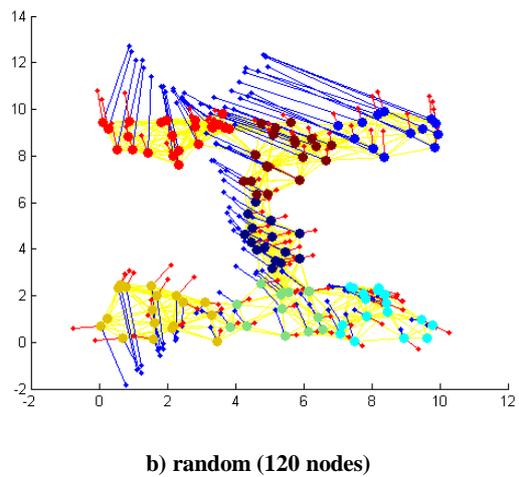

**b) random (120 nodes)**

**Figure 5. H-shaped topology (7 clusters, 4 anchors, R=1.8r)**

For each topology, we experimented with randomly placed nodes with a uniform distribution. Nodes are deployed on a 10r x 10r square, where r is a unit edge distance. For grid topology, nodes are not placed on the grid points, but 10%r placement error is added to each coordinate. This error is modeled as Gaussian noise. We changed the radio range R (1.3r, 1.5r, 1.8r, 2.0r, 2.5r) which led to a different connectivity of the network. For network clustering, we used kmeans function from Matlab. After computing the global map, different number of randomly chosen anchor nodes (3, 4, 6 0r 10) were used for aligning the relative positions to absolute positions [7]. We find the best linear transformation to generate the absolute map of the nodes.

Figure 6 to Figure 9 demonstrates the average performance of our algorithm as a function of connectivity for different topologies. Connectivity represents averages over 30 trials. Estimation errors are normalized with R, as proposed in [8][9]. As can be seen from the figures, CB-MDS performs smaller estimation error than MDS-MAP for both topologies.

Figure 6 shows that CB-MDS is better than MDS-MAP for all connectivity levels. It can also be mentioned that when the connectivity level is low, better results can be achieved if a small number of clusters are used. When connectivity level is 13, the best results are performed with 7 clusters. As the connectivity level increases, it is better to use more clusters. For connectivity level above 17 (20), the error is minimal if 10 (15) clusters are used to compute the local maps. This is true regardless of the number of anchors.

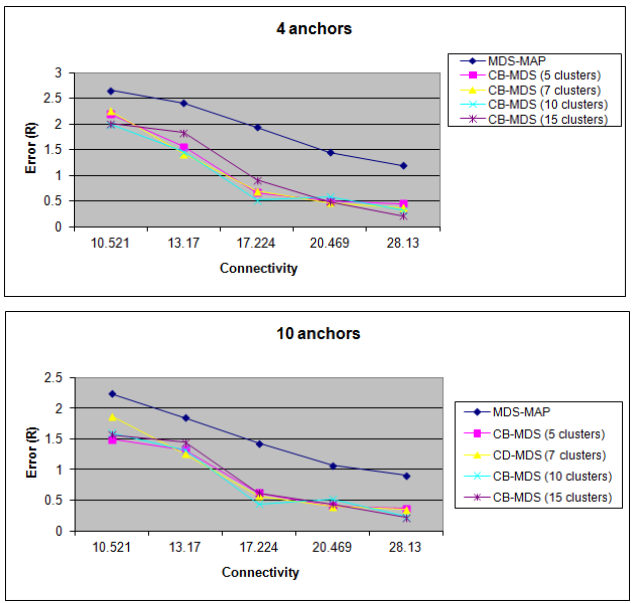

**Figure 6. Comparison of the MDS-MAP and CB-MDS on C-shaped topology (161 nodes randomly deployed)**

CB-MDS performs similar results for other topologies (see figure 7, Figure 8 and Figure 9).

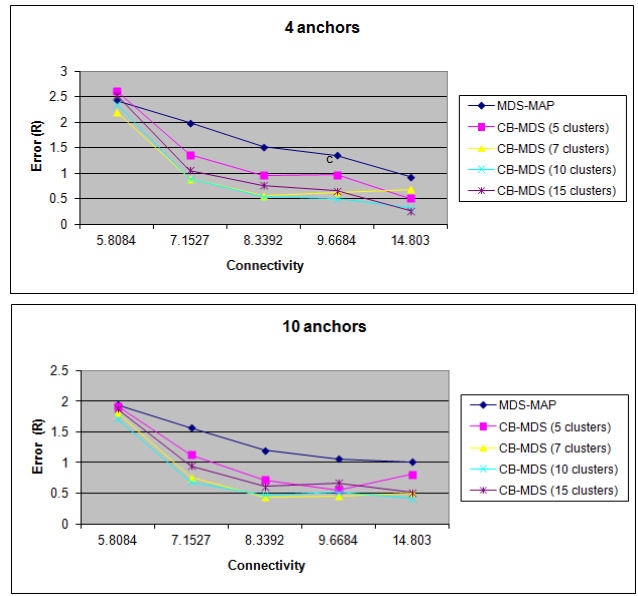

**Figure 7. Comparison of the MDS-MAP and Cluster-based MDS on a grid C-shaped topology (79 nodes randomly deployed)**

Number of anchors affects the results when the connectivity level is low. For high connectivity levels, there is no evident improvement (Figure 10).

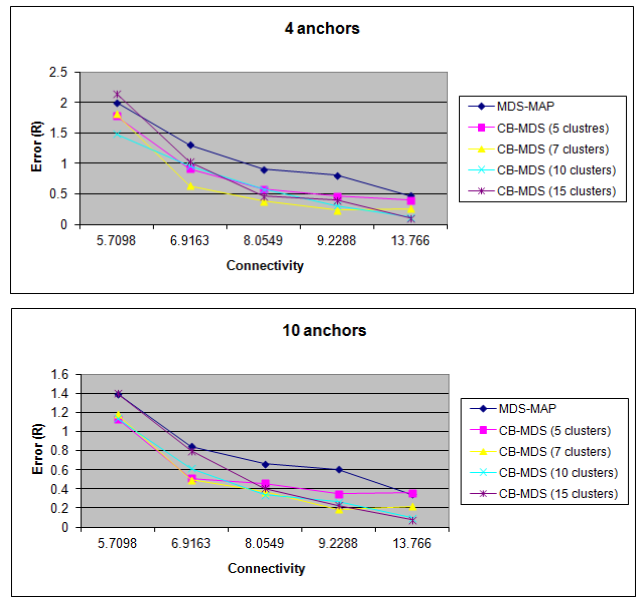

**Figure 8. Comparison of the MDS-MAP and Cluster-based MDS on a grid L-shaped topology (51 nodes randomly deployed)**

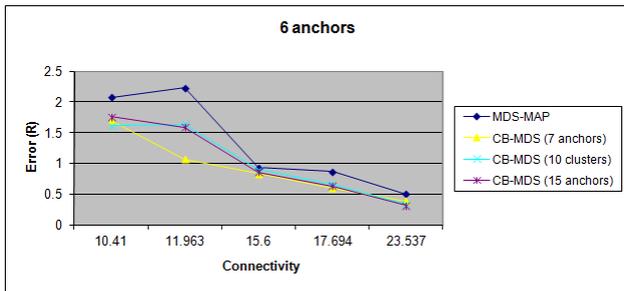

**Figure 9. Comparison of the MDS-MAP and CB-MDS on random H-shaped topology (110 nodes randomly deployed)**

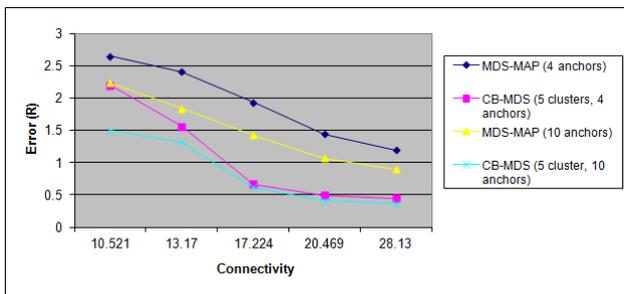

**Figure 10. The effect of number of anchors on the estimation error for MDS-MAP and CB-MDS on random C-shaped topology (110 nodes randomly deployed)**

## 5. CONCLUSION

In this paper, we presented a new cluster-based MDS algorithm for nodes localization in WSN. In our approach, network is divided into clusters and each cluster has one cluster-head for computing local information. Each cluster-head creates its own local relative map which consists of the nodes in its cluster. All local maps are merged into one global relative map using the best linear transformation. If anchor nodes are presented in the network, this global map can be transformed into global absolute map.

The results from our experiments (using the metric given in [8]), show better results than MDS-Map and MDS-MAP(P). Our algorithm estimates the nodes location with greater accuracy than MDS-MAP algorithm if applied on irregular topologies. If compared with MDS-MAP(P), our algorithm is less computational intensive, since in MDS-MAP(P) each node computes its local map. In CB-MDS, only cluster-heads do the computation.